\documentclass[12pt]{article}
\usepackage{amsmath,amssymb,amsfonts,color,graphicx,cite,color}
\usepackage{latexsym}
\usepackage{epsfig,psfrag,rotating,soul}
\usepackage{rotfloat}
\input paperdef

\evensidemargin \oddsidemargin
\allowdisplaybreaks

\begin{document}
\thispagestyle{empty}

\def\thefootnote{\fnsymbol{footnote}}

\begin{flushright}
\mbox{}
\end{flushright}

\vspace{0.5cm}

\begin{center}

\begin{large}
\textbf{Vacuum Stability Constraints on Flavour Mixing Parameters }
\\[2ex]
\textbf{ and Their Effect on Gluino Decays}
\end{large}

\vspace{1cm}

{\sc
S.~Ajmal$^{a,b}$%
\footnote{email: sehar.ajmal@pg.infn.it}
, M.~Rehman$^{a}$%
\footnote{email: m.rehman@comsats.edu.pk}%
, F.~Tahir$^{a}$ 
\footnote{email: farida.tahir@comsats.edu.pk}%
}

\vspace*{.7cm}
{\sl ${}^a$Department of Physics, Comsats University Islamabad, 44000 Islamabad, Pakistan}\\
{\sl ${}^b$Dipartimento di Fisica e Geologia, Università degli Studi di Perugia and INFN, Sezione di Perugia, Via A. Pascoli, I-06123, Perugia, Italy}\\

\end{center}

\vspace*{0.1cm}

\begin{abstract}
\noindent
In this article, we study the two body gluino decays 
$\tilde{g} \rightarrow \tilde{u}_i \bar{u}_g$, 
$\tilde{g} \rightarrow \tilde{d}_i \bar{d}_g$ with 
$i=1,2,...6$ and $g=1,2,3$ in the Minimal Supersymmetric Standard
Model (MSSM) with quark flavour violation (QFV). At the outset, constraints on QFV parameters $\delta
_{ij}^{U/D(LR)}$ and $\delta _{ij}^{U/D(RL)}$ with $i,j=1,2,3$ and $(i\neq j)
$ are calculated for a specific set of MSSM parameters using charge and color
breaking minima (CCB) and unbounded from below minima (UFB) conditions. These
constraints are more stringent compared to the constraints coming from B-physics observables (BPO), that are already available in
literature. In the second step, we re-calculate the partial decay widths of $\tilde{g} \rightarrow \tilde{u}_i \bar{u}_g$, 
$\tilde{g} \rightarrow \tilde{d}_i \bar{d}_g$ for
the allowed range of QFV parameters. The partial decay widths can reach upto 120 GeV in the RR sector of the squarks mass matrices while in the LL sector it can be $\approx 90 \gev$. The QFV in the LR/RL sector is usually ignored due to stringent CCB and UFB constraints. However, our analysis reveals that this mixing can contribute upto $\approx 12 \gev$ for some parameter points and should not be ignored. We hope that these results will prove helpful for the experimental searches of gluinos at the current and future colliders.

\end{abstract}

\def\thefootnote{\arabic{footnote}}
\setcounter{page}{0}
\setcounter{footnote}{0}

\newpage



\section{Introduction}

Minimal Supersymmetric Standard Model (MSSM) \cite{Nilles:1983ge, Haber:1984rc, Barbieri:1987xf} is a rather
enthralling theory Beyond the Standard Model (SM)\cite{Glashow:1961tr,Weinberg:1967tq,Goldstone:1962es}. It is
able to answer many of the queries that remained unanswered in SM. Search
for the supersymmetric particles is one of the major tasks at the large
hadron collider experiment at CERN. Therefore, a lot of attention has been
devoted to the study of sparticle decays. Gluino decays are particularly
interesting as the bounds on the gluino mass are progressively increasing\cite{Aad:2020nyj}.
Analyses of these decays provide the new techniques for experimental
searches and deep insight of couplings between standard particles and their
super-partners.

Contrary to SM, where CKM matrix is the only source of quark flavour
violation (QFV), MSSM contains extra sources~\cite{AranaCatania:2012sn, Arana-Catania:2013nha, Gomez:2014uha}, 
in the form of quark-squarks misalignments which appear as the non-diagonal
entries in the squarks mass matrices and is parametrized in terms of flavor
violating (FV) deltas $\delta _{ij}^{FAB}$, where $i\neq j$ $(i,j=1,2,3)$
and $F=Q,U,D;$ $A,B=L,R$. These FV deltas can result in the large amplitudes for flavour changing
neutral current (FCNC) processes. Therefore, there are stringent constraints on these deltas
due to FCNC processes. It has been shown in some recent studies 
\cite{Gomez:2015ila, Gomez:2015duj} that the flavor violating deltas may result in interesting
phenomenological effects, contrary to the some previous studies where these deltas were put to zero by hand.

As has already been scientifically broadcasted, the FV delta can also impact
the gluino decays. For example gluino decays into quark and squark at tree
(with QFV) as well as at loop (without QFV) level were analysed in \cite%
{Beenakker:1996dw, Choi:2008pi, Kramer:2009kp, Hurth:2009ke, Bartl:2009au, Bartl:2011wq, Heinemeyer:2011ab}. 
Furthermore, QFV gluino decays into a quark and squark at one loop level (mainly focused on $RR$ mixing) 
were investigated in \cite{Eberl:2017pbu} considering the strong suppression of the $LL$, $LR$ and $RL$ mixings due to
FCNC, charge and color breaking minima (CCB) and unbounded from below minima
(UFB) \cite{Casas:1996de}.

We have extended this analysis by incorporating the contribution of LL, LR\
and RL mixing in two body gluino decays. As mentioned earlier, the
indirect bounds on $\delta _{ij}^{QAB}$ coming from FCNC processes are very
strong. First and second generation mixing is severely constrained by $K$%
-Physics. However, there is some room for the second and third generation
mixing. The constraints on $\delta _{23}^{QAB}$ mainly come from B-Physics
Observables (BPO). These constraints for a specific set of parameters were
calculated in \cite{Arana-Catania:2014ooa}. Using the same set of parameters (with a slight
modification of $M_{A}$\ value), we have extended their analysis and
calculated the CCB and UFB bounds on these deltas. The vacuum stability (CCB
and UFB) conditions turned out to be more constraining in the $LR,$ $RL$\ sector
compared to the constraints from BPO. As a final step, we have analyzed the
two body gluino decay using these deltas. For numerical analysis, we have
used \fa/\fc~setup~\cite{Hahn:2000kx, Hahn:1998yk}.

The paper is organized as follows: in \refse{sec:model_MSSM} we discuss
the salient features of the MSSM and set the notation. \refse{sec:CompSetup}
is dedicated to the analytical calculation of gluino decay and some details
of BPO, CCB and UFB. In \refse{sec:NResults} we discuss our numerical
analysis followed by the conclusions of this work in \refse{sec:conclusions}.

\section{Model set-up}
\label{sec:model_MSSM}
 \textbf{The R-parity conserving superpotential $W$ of MSSM}
is given as~\cite{Nilles:1983ge, Haber:1984rc, Barbieri:1987xf}%
\begin{equation}
W=\epsilon _{\alpha \beta }\left[ (\lambda _{l})_{ij}\hat{H}_{1}^{\alpha }\hat{%
L}_{\iota }^{\beta }\hat{E}_{j}^{C}+(\lambda _{d})_{ij}\hat{H}_{1}^{\alpha }%
\hat{Q}_{\iota }^{\beta }\hat{D}_{j}^{C}+(\lambda _{u})_{ij}\hat{H}%
_{2}^{\alpha }\hat{Q}_{\iota }^{\beta }\hat{U}_{j}^{C}-\mu \hat{H}%
_{1}^{\alpha }\hat{H}_{2}^{\beta }\right]   \notag
\end{equation}%
where $\lambda $ represents the Yukawa couplings and $\mu$ is the Higgs
mass parameter. $\hat{Q}$ is for left-handed quark and squark doublet, $\hat{%
U}$ is for right-handed up-type quark and squark singlet$,$ and $\hat{D}$ is
for right-handed down--type quark and squark singlet. For leptonic fields,
there are $\hat{L}$ for left-handed lepton and slepton doublet and $\hat{E}$
for right handed lepton and slepton singlet. There are no right handed
neutrinos present in MSSM. $\hat{H}_{1}$ and $\hat{H}_{2}$ are the 2 Higgs
superfields. As SUSY is not exact symmetry therefore we have to incorporate
SUSY breaking in the MSSM. Accordingly, the soft-SUSY-Breaking Lagrangian
can be written as:%
\begin{align}
-%
\mathcal{L}%
_{soft}  & =(m_{\tilde{Q}}^{2})_{i}^{j}\tilde{Q}^{\dagger\iota}\tilde{Q}%
_{j}+(m_{\tilde{U}}^{2})_{j}^{i}\tilde{U}_{i}^{\ast}\tilde{U}^{j}%
+(m_{\tilde{D}}^{2})_{j}^{i}\tilde{D}_{i}^{\ast}\tilde{D}^{j}+(m_{\tilde{L}%
}^{2})_{i}^{j}\tilde{L}^{\dagger\iota}\tilde{L}_{j}\nonumber\\
& +(m_{\tilde{E}}^{2})_{j}^{i}\tilde{E}_{\iota}^{\ast}\tilde{E}^{j}+m_{H_{1}%
}^{2}\hat{H}_{1}^{\dagger}\hat{H}_{1}+m_{H_{2}}^{2}\hat{H}_{2}^{\dagger}%
\hat{H}_{2}+\left(  B\mu\hat{H}_{1}^{\dagger}\hat{H}_{2}+h.c\right)
\nonumber\\
& +(({\cal A}^{d})_{ij}\hat{H}_{1}\tilde{D}_{i}^{\ast}\tilde{Q}_{j}+(
{\cal A}^{u})_{ij}\hat{H}_{2}\tilde{U}_{i}^{\ast}\tilde{Q}_{j}+({\cal A}^{l}%
)_{ij}\hat{H}_{1}\tilde{E}_{\iota}^{\ast}\tilde{E}_{j}\nonumber\\
& +\frac{1}{2}M_{1}\tilde{B}_{L}^{0}\tilde{B}_{L}^{0}+\frac{1}{2}M_{2}%
\tilde{W}_{L}^{a}\tilde{W}_{L}^{a}+\frac{1}{2}M_{3}\tilde{G}^{a}\tilde{G}%
^{a}+h.c)\label{soft susy}%
\end{align}
$m_{\hat{Q}}^{2},$ $m_{\hat{U}}^{2},$ $m_{\tilde{D}}^{2}$ correspond to a $%
3\times 3$ mass matrices in family space for the soft masses of left and
right handed squarks, whereas, left and right handed sleptons mass matrices
are given by $m_{\tilde{L}}^{2}$ and $m_{\tilde{E}}^{2}.$ $m_{H_{1}}^{2}$
and $m_{H_{2}}^{2}$ contain soft masses of Higgs sector. ${\cal A}^{u},$ $%
{\cal A}^{d},$ ${\cal A}^{l}$ are the $3\times 3$ matrices of up-type,
down-type and charged lepton trilinear couplings, respectively. At last $%
M_{1},$ $M_{2},$ $M_{3}$ are the bino, wino, and gluino mass terms,
respectively. Sfermions gain mass after the EW symmetry breaking. F, D-type
and soft SUSY breaking terms appear in the mass matrix of sfermions. Mass
matrix of sfermions is written as 
\begin{equation}
\cM_{\tilde q}^2 =\left( \begin{array}{cc}
M^2_{\tilde q \, LL} & M^2_{\tilde q \, LR} \\[.3em] 
M_{\tilde q \, LR}^{2 \, \dagger} & M^2_{\tilde q \,RR}
\end{array} \right), \qquad \tilde q= \tilde u, \tilde d~,
\label{eq:blocks-matrix}
\end{equation} 
 where:
 \begin{alignat}{5}
 M_{\tilde u \, LL \, ij}^2 
  = &  m_{\tilde U_L \, ij}^2 + \left( m_{u_i}^2
     + (T_3^u-Q_u\sw^2 ) M_Z^2 \cos 2\beta \right) \delta_{ij},  \notag\\
 M^2_{\tilde u \, RR \, ij}
  = &  m_{\tilde U_R \, ij}^2 + \left( m_{u_i}^2
     + Q_u\sw^2 M_Z^2 \cos 2\beta \right) \delta_{ij} \notag, \\
  M^2_{\tilde u \, LR \, ij}
  = &  \nu_2 {\cal A}_{ij}^u- m_{u_{i}} \mu \cot \beta \, \delta_{ij},
 \notag, \\
 M_{\tilde d \, LL \, ij}^2 
  = &  m_{\tilde D_L \, ij}^2 + \left( m_{d_i}^2
     + (T_3^d-Q_d \sw^2 ) M_Z^2 \cos 2\beta \right) \delta_{ij},  \notag\\
 M^2_{\tilde d \, RR \, ij}
  = &  m_{\tilde D_R \, ij}^2 + \left( m_{d_i}^2
     + Q_d\sw^2 M_Z^2 \cos 2\beta \right) \delta_{ij} \notag, \\
  M^2_{\tilde d \, LR \, ij}
  = &  \nu_1 {\cal A}_{ij}^d- m_{d_{i}} \mu \tb \, \delta_{ij}~,
\label{eq:SCKM-entries}
\end{alignat}
with, $i,j=1,2,3$, $Q_u=2/3$, $Q_d=-1/3$, $T_3^u=1/2$ and
$T_3^d=-1/2$. $(m_{u_1},m_{u_2}, m_{u_3})=(m_u,m_c,m_t)$, $(m_{d_1},m_{d_2},
m_{d_3})=(m_d,m_s,m_b)$ are the quark masses and $(m_{l_1},m_{l_2},
m_{l_3})=(m_e,m_\mu,m_\tau)$ are the lepton masses.
In the SM, flavour states can mix with each other through CKM matrix and
diagonalization of CKM gives physical mass eigenstates. Accordingly, the 6
component flavour eigenstates of squarks can mix together via $6\times 6$
rotation matrix. Rotation from interaction eigenstates to mass eigenstates
is performed as, 
\begin{equation}
\left( 
\begin{array}{c}
\tilde{u}_{1} \\ 
\tilde{u}_{2} \\ 
\tilde{u}_{3} \\ 
\tilde{u}_{4} \\ 
\tilde{u}_{5} \\ 
\tilde{u}_{6}%
\end{array}%
\right) =R^{\tilde{u}}\left( 
\begin{array}{c}
\tilde{u}_{L} \\ 
\tilde{c}_{L} \\ 
\tilde{t}_{L} \\ 
\tilde{u}_{R} \\ 
\tilde{c}_{R} \\ 
\tilde{t}_{R}%
\end{array}%
\right) ,\text{ \  \ }\left( 
\begin{array}{c}
\tilde{d}_{1} \\ 
\tilde{d}_{2} \\ 
\tilde{d}_{3} \\ 
\tilde{d}_{4} \\ 
\tilde{d}_{5} \\ 
\tilde{d}_{6}%
\end{array}%
\right) =R^{\tilde{d}}\left( 
\begin{array}{c}
\tilde{d}_{L} \\ 
\tilde{s}_{L} \\ 
\tilde{b}_{L} \\ 
\tilde{d}_{R} \\ 
\tilde{s}_{R} \\ 
\tilde{b}_{R}%
\end{array}%
\right)   \label{6x6 rotation}
\end{equation}%
here $\tilde{u}_{1},...$ are up-type squarks in physical basis, $R^{\tilde{u}%
}$ is the corresponding rotation matrix and $\tilde{u}_{L},\tilde{c}_{L},%
\tilde{t}_{L},\tilde{u}_{R},\tilde{c}_{R},\tilde{t}_{R}$ are up-type squarks
interaction eigenstates. Likewise,$\  \tilde{d}_{1},...$ are down-type
squarks in physical basis, $R^{\tilde{d}}$ is the corresponding rotation
matrix and $\tilde{d}_{L},\tilde{s}_{L},\tilde{b}_{L},\tilde{d}_{R},\tilde{s}%
_{R},\tilde{b}_{R}$ are down-type squarks interaction eigenstates.
Flavour mixing parameters are given by $\delta _{ij}^{FAB}$ in off-diagonal
entries in mass squared matrix as well as trilinear coupling matrices, where 
$F=Q,U,D.$ $A,B=L,R$ and $i,j=1,2,3$ with $i\neq j$. In flavour space $M_{%
\tilde{f}}^{2}$ in $3\times 3$ block form for squarks are given as%
\begin{equation}
m_{\tilde{U}_{L}}^{2}=\left( 
\begin{array}{ccc}
m_{\tilde{Q}_{1}}^{2} & \delta _{12}^{QLL}m_{\tilde{Q}_{1}}^{2}m_{\tilde{Q}%
_{2}}^{2} & \delta _{13}^{QLL}m_{\tilde{Q}_{1}}^{2}m_{\tilde{Q}_{3}}^{2} \\ 
\delta _{21}^{QLL}m_{\tilde{Q}_{2}}^{2}m_{\tilde{Q}_{1}}^{2} & m_{\tilde{Q}%
_{2}}^{2} & \delta _{23}^{QLL}m_{\tilde{Q}_{2}}^{2}m_{\tilde{Q}_{3}}^{2} \\ 
\delta _{31}^{QLL}m_{\tilde{Q}_{3}}^{2}m_{\tilde{Q}_{1}}^{2} & \delta
_{32}^{QLL}m_{\tilde{Q}_{3}}^{2}m_{\tilde{Q}_{2}}^{2} & m_{\tilde{Q}_{3}}^{2}%
\end{array}%
\right)   \label{Mass1}
\end{equation}%
\begin{equation}
m_{\tilde{D}_{L}}^{2}=V_{CKM}^{\dagger }m_{\tilde{U}_{L}}^{2}V_{CKM}
\label{Mass2}
\end{equation}%
\begin{equation}
m_{\tilde{U}_{R}}^{2}=\left( 
\begin{array}{ccc}
m_{\tilde{U}_{1}}^{2} & \delta _{12}^{URR}m_{\tilde{U}_{1}}^{2}m_{\tilde{U}%
_{2}}^{2} & \delta _{13}^{URR}m_{\tilde{U}_{1}}^{2}m_{\tilde{U}_{3}}^{2} \\ 
\delta _{21}^{URR}m_{\tilde{U}_{2}}^{2}m_{\tilde{U}_{1}}^{2} & m_{\tilde{U}%
_{2}}^{2} & \delta _{23}^{URR}m_{\tilde{U}_{2}}^{2}m_{\tilde{U}_{3}}^{2} \\ 
\delta _{31}^{URR}m_{\tilde{U}_{3}}^{2}m_{\tilde{U}_{1}}^{2} & \delta
_{32}^{URR}m_{\tilde{U}_{3}}^{2}m_{\tilde{U}_{2}}^{2} & m_{\tilde{U}_{3}}^{2}%
\end{array}%
\right)   \label{Mass3}
\end{equation}%
\begin{equation}
m_{\tilde{D}_{R}}^{2}=\left( 
\begin{array}{ccc}
m_{\tilde{D}_{1}}^{2} & \delta _{12}^{DRR}m_{\tilde{D}_{1}}^{2}m_{\tilde{D}%
_{2}}^{2} & \delta _{13}^{DRR}m_{\tilde{D}_{1}}^{2}m_{\tilde{D}_{3}}^{2} \\ 
\delta _{21}^{DRR}m_{\tilde{D}_{2}}^{2}m_{\tilde{D}_{1}}^{2} & m_{\tilde{D}%
_{2}}^{2} & \delta _{23}^{DRR}m_{\tilde{D}_{2}}^{2}m_{\tilde{D}_{3}}^{2} \\ 
\delta _{31}^{DRR}m_{\tilde{D}_{3}}^{2}m_{\tilde{D}_{1}}^{2} & \delta
_{32}^{DRR}m_{\tilde{D}_{3}}^{2}m_{\tilde{D}_{2}}^{2} & m_{\tilde{D}_{3}}^{2}%
\end{array}%
\right)   \label{Mass4}
\end{equation}%
similarly, for trilinear coupling we have 
\begin{equation}
\nu_{2} {\cal A}^{u}=\left( 
\begin{array}{ccc}
m_{u}A_{u} & \delta _{12}^{ULR}m_{\tilde{Q}_{1}}^{2}m_{\tilde{U}_{2}}^{2}
& \delta _{13}^{ULR}m_{\tilde{Q}_{1}}^{2}m_{\tilde{U}_{3}}^{2} \\ 
\delta _{21}^{ULR}m_{\tilde{Q}_{2}}^{2}m_{\tilde{U}_{1}}^{2} & m_{c}A_{c}
& \delta _{23}^{ULR}m_{\tilde{Q}_{2}}^{2}m_{\tilde{U}_{3}}^{2} \\ 
\delta _{31}^{ULR}m_{\tilde{Q}_{3}}^{2}m_{\tilde{U}_{1}}^{2} & \delta
_{32}^{ULR}m_{\tilde{Q}_{3}}^{2}m_{\tilde{U}_{2}}^{2} & m_{t}A_{t}%
\end{array}%
\right)   \label{tri}
\end{equation}%

\begin{equation}
\nu _{1} {\cal A}^{d}=\left( 
\begin{array}{ccc}
m_{d}A_{d} & \delta _{12}^{DLR}m_{\tilde{Q}_{1}}^{2}m_{\tilde{U}_{2}}^{2}
& \delta _{13}^{DLR}m_{\tilde{Q}_{1}}^{2}m_{\tilde{U}_{3}}^{2} \\ 
\delta _{21}^{DLR}m_{\tilde{Q}_{2}}^{2}m_{\tilde{U}_{1}}^{2} & m_{s}A_{s}
& \delta _{23}^{DLR}m_{\tilde{Q}_{2}}^{2}m_{\tilde{U}_{3}}^{2} \\ 
\delta _{31}^{DLR}m_{\tilde{Q}_{3}}^{2}m_{\tilde{U}_{1}}^{2} & \delta
_{32}^{DLR}m_{\tilde{Q}_{3}}^{2}m_{\tilde{U}_{2}}^{2} & m_{b}A_{b}%
\end{array}%
\right)   \label{tri}
\end{equation}%

As discussed in introduction, the mixing between 2nd and 3rd generation is
very important. So, the dimensionless parameters $(\delta _{ij}^{F})^{AB}$ for second and third generation mixing are
encoded as%
\begin{eqnarray}
\delta _{23}^{QLL} &\equiv &\frac{m_{\tilde{Q}23}^{2}}{\sqrt{m_{\tilde{Q}%
22}^{2}m_{\tilde{Q}33}^{2}}}\sim \tilde{c}_{L}-\tilde{t}_{L}\text{ }mixing
\label{DeltaQ} \\
\delta _{23}^{URR} &\equiv &\frac{m_{\tilde{U}23}^{2}}{\sqrt{m_{\tilde{U}%
22}^{2}m_{\tilde{U}33}^{2}}}\sim \tilde{c}_{R}-\tilde{t}_{R}\text{ }mixing
\label{DeltaU} \\
\delta _{23}^{URL} &\equiv & \frac{\nu _{2} {\cal A}^{%
u}_{23}}{\sqrt{m_{\tilde{U}22}^{2}m_{\tilde{Q}33}^{2}}}\sim \tilde{c}%
_{R}-\tilde{t}_{L}\text{ }mixing  \label{trilinearURL} \\
\delta _{23}^{ULR} &\equiv & \frac{\nu _{2} {\cal A}^{%
u}_{32}}{\sqrt{m_{\tilde{U}22}^{2}m_{\tilde{Q}33}^{2}}}\sim \tilde{c}%
_{L}-\tilde{t}_{R}\text{ }mixing  \label{trilinearULR}
\end{eqnarray}%
on the same footing one can write relations for down type mixing as%
\begin{eqnarray}
\delta _{23}^{DRR} &\equiv &\frac{m_{\tilde{D}23}^{2}}{\sqrt{m_{\tilde{D}%
22}^{2}m_{\tilde{D}33}^{2}}}\sim \tilde{s}_{R}-\tilde{b}_{R}\text{ }mixing
\label{deltaD} \\
\delta _{23}^{DLR} &\equiv &\frac{\nu _{1}{\cal A}^{d}
_{32}}{\sqrt{m_{\tilde{D}22}^{2}m_{\tilde{Q}33}^{2}}}\sim \tilde{s}%
_{L}-\tilde{b}_{R}\text{ }mixing  \label{trilinearDLR} \\
\delta _{23}^{DRL} &\equiv & \frac{\nu _{1} {\cal A}^{%
d}_{23}}{\sqrt{m_{\tilde{D}22}^{2}m_{\tilde{Q}33}^{2}}}\sim \tilde{s}%
_{R}-\tilde{b}_{L}\text{ }mixing  \label{trilinearDRL}
\end{eqnarray}%
these relations will be helpful in the study of QFV gluino decays.

\section{Computational setup}
\label{sec:CompSetup}
\subsection{Two body Gluino Decays}
Gluino $\tilde{g},$ the MSSM partner of gluon can only decay via squarks,
either on-shell or off-shell. The decay pattern is given as%
\begin{equation*}
\tilde{g}\rightarrow \tilde{u}_{i}\bar{u}_{g},\text{ \  \  \  \  \  \  \ }\tilde{g}%
\rightarrow \tilde{d}_{i}\bar{d}_{g}
\end{equation*}%
with $i=1,2,\cdots ,6$ and $g=1,2,3,$ these decay patterns are dominant
because of QCD strength. The interaction Lagrangian of gluino-quark-squark
is given as ~\cite{Eberl:2017pbu}%
\begin{align*}
\mathcal{L}%
_{\tilde{g}\tilde{q}_{i}\bar{q}_{g}}  & =-\sqrt{2}g_{s}\lambda^{a}%
[\overline{\tilde{g}}^{a}(R_{i,g}^{\tilde{q}}P_{L}-R_{i,g+3}^{\tilde{q}}%
P_{R})q_{g}\tilde{q}_{i}^{\ast}\\
&+\overline{q}_{g}(R_{i,g}^{\tilde{q}^{\ast}}P_{R}-R_{i,g+3}^{\tilde{q}%
^{\ast}}P_{L})\tilde{g}^{a}\tilde{q}_{i}]
\end{align*}
here $\lambda ^{a}$ is the generator of SU(3), $R^{\tilde{q}}$ are the
rotation matrices for squarks, and $g_{s}$ is the QCD strength. Tree level partial decay width of \ $\tilde{g}%
\rightarrow \tilde{u}_{i}\bar{u}_{g}$ is written as%
\begin{align}
\Gamma (\tilde{g}\rightarrow \tilde{u}_{i}\bar{u}_{g})&=\frac{C\mathbf{P}(%
m_{0}^{2},m_{1}^{2},m_{2}^{2})}{32\pi m_{0}^{3}} \label{decay-width}  \\
&\left[(m_{0}^{2}-m_{1}^{2}+m_{2}^{2})(\left \vert \alpha _{L}\right \vert
^{2}+\left \vert \alpha _{R}\right \vert ^{2}) 
+2m_{0}m_{2}(\alpha
_{L}^{\ast }\alpha _{R}+\alpha _{L}\alpha _{R}^{\ast })\right] \nonumber
\end{align}
where C is the color factor which is equal to 1/8, $m_{0}=m_{\tilde{g}},$ $%
m_{1}=m_{\tilde{u}_{i}}$ and $m_{2}=m_{u_{g}}$, $\alpha _{L,R}$ is given as%
\begin{equation*}
\alpha _{L}=-\sqrt{2}g_{s}\lambda ^{a}R_{i,g}^{\tilde{u}\text{ }}\text{ , \
\  \ }\alpha _{R}=\sqrt{2}g_{s}\lambda ^{a}R_{i,g+3}^{\tilde{u}\text{ }}
\end{equation*}%
and $\mathbf{P}(m_{0}^{2},m_{1}^{2},m_{2}^{2})$ is the triangular function defined as 
\begin{equation*}
\mathbf{P}(m_{0}^{2},m_{1}^{2},m_{2}^{2})=\frac{1}{2m_{0}}\sqrt{
m_{0}^{4}+m_{1}^{4}+m_{2}^{4}-2m_{0}^{2}m_{1}^{2}-2m_{0}^{2}m_{2}^{2}-2m_{1}^{2}m_{2}^{2}}
\end{equation*}
For the decay width $\tilde{g}\rightarrow \tilde{d}_{i}\bar{d}_{g}$,  we can use
\refeq{decay-width} by putting $m_{1}=m_{\tilde{d}_{i}}$, $m_{2}=m_{d_{g}}$ and replacing $R_{i,g}^{\tilde{u}}$ with the corresponding rotation matrix in the down sector i.e. $R_{i,g}^{\tilde{d}}$.
\subsection{Constraints on $\protect \delta^{FAB}_{ij}$}

Flavour violating deltas $\delta _{ij}^{FAB}$ in the squark sector can be
constrained using electroweak precision observables (EWPO), BPO, CCB and
UFB. For the specific set of parameter points that we will use
in this work, the constraints from BPO have already been calculated in \cite%
{Arana-Catania:2014ooa}. However, we calculate the constraints from CCB and UFB conditions \cite%
{Casas:1996de} that are relevant for $LR$ and $RL$ sectors. Details about this
calculation are presented below.

\subsubsection{CCB and UFB bounds}

CCB minima appear whenever color and electrically charged particles gain VEV
which violates the exact symmetry of $SU(3)_{c}\times U(1)_{Y}$. On the
other hand, to make sure that potential is bounded from below, UFB
constraints are needed. These two (CCB\ and UFB) are named as vacuum
stability bounds and they dictate stronger constraints on $(A_{f})_{ij}$ than
those imposed by the FCNC. Here we are incorporating charged and colored
fields in symmetry breaking Lagrangian, as the scalar potential of MSSM
contains sfermions scalar fields. Following the approach of \cite{Casas:1996de}, we can write the off-diagonal term $({\cal A}^{u})_{ij}$ for $i=1$ and $j=2$ as
\begin{equation}
({\cal A}^{u})_{12}\tilde{u}_{L}H_{2}^{0}\tilde{c_{R}^{\ast}}+h.c
\label{6.1}
\end{equation}%
by adding these contributions, scalar potential is extended and extra terms
in scalar potential lead to CCB minima if 
\begin{equation*}
\left \vert ({\cal A}^{u})_{12}\right \vert ^{2}\leq \lambda _{c}^{2}\left( m_{%
\tilde{u}_{L}}^{2}+m_{\tilde{c}_{R}}^{2}+m_{2}^{2}\right) 
\end{equation*}%
In order to get true ground state, one has to satisfy these constraints.
Now, we can easily generalize this bound for up and down squarks as%
\begin{align}
\left \vert ({\cal A}^{u})_{ij}\right \vert ^{2}& \leq \lambda
_{u_{k}}^{2}\left( m_{\tilde{u}_{L_{i}}}^{2}+m_{\tilde{u}%
_{R_{j}}}^{2}+m_{2}^{2}\right) ,\text{ \  \  \  \  \  \  \  \  \ }k=Max(i,j)
\label{6.3} \\
\left \vert ({\cal A}^{d})_{ij}\right \vert ^{2}& \leq \lambda
_{d_{k}}^{2}\left( m_{\tilde{d}_{L_{i}}}^{2}+m_{\tilde{d}%
_{Rj}}^{2}+m_{1}^{2}\right) ,\text{ \  \  \  \  \  \  \  \  \ }k=Max(i,j)
\label{6.4}
\end{align}

One can also modify flavour violating deltas given by \refeqs
{trilinearULR} and \ref{trilinearDLR}) (for $i,j$) as%
\begin{align}
\delta _{ij}^{ULR}& \leq M_{u_{k}}\frac{\left[ 2M_{av}^{2(\tilde{u}%
)}+m_{2}^{2}\right] ^{1/2}}{M_{av}^{2(\tilde{u})}},\text{ \  \  \  \  \  \ }%
k=Max(i,j)  
\label{stability1} \\
\delta _{ij}^{DLR}& \leq M_{d_{k}}\frac{\left[ 2M_{av}^{2(\tilde{d}%
)}+m_{1}^{2}\right] ^{1/2}}{M_{av}^{2(\tilde{d})}},\text{ \  \  \  \  \  \ }%
k=Max(i,j)  
\label{stability2}
\end{align}%
where $M$ is the mass of quarks, and $M_{av}$ is the average mass of
squarks. Also, $m_{1}^{2}$ and $m_{2}^{2}$ are given as%
\begin{align}
m_{1}^{2}& =(M_{A}^{2}+M_{Z}^{2})\sin ^{2}\beta -\frac{1}{2}M_{Z}^{2}  \notag
\\
m_{2}^{2}& =(M_{A}^{2}+M_{Z}^{2})\cos ^{2}\beta -\frac{1}{2}M_{Z}^{2}
\label{modified M}
\end{align}
$M_{A}$ and $M_{Z}$ is the mass of CP-odd higgs boson and Z boson,
respectively.

Correspondingly, UFB bounds can be calculated by using \refeq{6.1},
additional fields (sneutrinos) are also included as compared to CCB.
Firstly, $({\cal A}^{u})_{ij}$ will be chosen, all possible contributions are
taken into account. The scalar potential becomes negative except if 
\begin{equation}
\left \vert ({\cal A}^{u})_{ij}\right \vert ^{2}\leq \lambda _{u_{k}}^{2}\left(
m_{\tilde{u}_{L_{i}}}^{2}+m_{\tilde{u}_{R_{j}}}^{2}+m_{\tilde{e}%
_{L_{p}}}^{2}+m_{\tilde{e}_{R_{q}}}^{2}\right) ,\text{ \  \  \  \ }k=Max(i,j),%
\text{ \ }p\neq q\text{\  \  \ }  \label{trilinear1}
\end{equation}%
similarly, for down type we have 
\begin{equation}
\left \vert ({\cal A}^{d})_{ij}\right \vert ^{2}\leq \lambda _{d_{k}}^{2}\left(
m_{\widetilde{d}_{L_{i}}}^{2}+m_{\tilde{d}_{R_{j}}}^{2}+m_{\tilde{\nu}%
}^{2}\right) ,\text{ \  \  \  \ }k=Max(i,j)  \label{trilinear2}
\end{equation}%
Now, we can modify flavour violating deltas given by \refeqs{trilinearULR}
and (\ref{trilinearDLR}) (for $i,j$) as%
\begin{equation}
\delta _{ij}^{ULR}\leq M_{u_{k}}\frac{\left[ 2M_{av}^{2(\tilde{u}%
)}+2M_{av}^{2(\tilde{l})}\right] ^{1/2}}{M_{av}^{2(\tilde{u})}},\text{ \  \  \
\  \  \ }k=Max(i,j)  
\label{stability3}
\end{equation}%
\begin{equation}
\delta _{ij}^{DLR}\leq M_{d_{k}}\frac{\left[ 2M_{av}^{2(\tilde{d}%
)}+M_{av}^{2(\tilde{l})}\right] ^{1/2}}{M_{av}^{2(\tilde{d})}},\text{ \  \  \
\  \  \ }k=Max(i,j)  
\label{stability4}
\end{equation}

\section{Numerical Results}

\label{sec:NResults}

In this section, we will present our numerical results for the partial decay
width of $\widetilde{g}\rightarrow \tilde{u}_{i}\bar{u}_{g}$ and $\widetilde{%
g}\rightarrow \tilde{d}_{i}\bar{d}_{g}$ ($i=1,2,\cdots ,6$ and $g=1,2,3)$
a set of parameter points taken from~\cite{Arana-Catania:2014ooa}.
However, we have assigned \cp-odd Higgs mass $M_{A}$ higher value
to make these points consistent with the present experimental results from
LHC.

For simplicity, and to reduce the number of independent MSSM input
parameters, we assume degenerated soft masses for the chiral squarks\ and
sleptons of first and second generations. Throughout this analysis equal
trilinear couplings are chosen for the stop and sbottom (3rd generation)
squarks as well as for the sleptons, whereas the trilinear couplings for the
1st and 2nd generations are ignored. Furthermore, we assume an approximate GUT relation for the gaugino
soft-SUSY-breaking parameters. The pseudoscalar Higgs mass $\MA$ and the $%
\mu $ are taken as independent input parameters. In summary, the five points
S1\dots S5 are defined in terms of the following subset of ten input MSSM
parameters: 
\begin{align*}
& m_{\tilde{L}_{1}}=m_{\tilde{L}_{2}}\,, & & m_{\tilde{L}_{3}}\,, & & (\text{%
with~}m_{\tilde{L}_{i}}=m_{\tilde{E}_{i}},\ i=1,2,3) \\
& m_{\tilde{Q}_{1}}=m_{\tilde{Q}_{2}}, & & m_{\tilde{Q}_{3}}\,, & & (\text{%
with~}m_{\tilde{Q}_{i}}=m_{\tilde{U}_{i}}=m_{\tilde{D}_{i}},\ i=1,2,3) \\
& A_{t}=A_{b}\,, & & A_{\tau}\,, & & \\
& M_{2}=2M_{1}=M_{3}/4\,, & & \mu \,, & & \\
& \MA \,, & & \tb \,. & &
\end{align*}

\begin{table}[h!]
\centerline{\begin{tabular}{|c|c|c|c|c|c|}
\hline
 & S1 & S2 & S3 & S4 & S5 \\\hline
$m_{\tilde L_{1,2}}$& 500 & 750 & 1000 & 500 & 800  \\
$m_{\tilde L_{3}}$ & 500 & 750 & 1000 & 500 & 500  \\
$M_2$ & 500 & 500 & 500 & 750 & 500 \\
$A_{\tau}$ & 500 & 750 & 1000 & 0 & 500 \\
$\mu$ & 400 & 400 & 400 & 800 & 400  \\
$\tb$ & 20 & 30 & 50 & 10 & 40 \\
$\MA$ & 1300 & 1500 & 1800 & 1000 & 1700 \\
$m_{\tilde Q_{1,2}}$ & 2000 & 2000 & 2000 & 2500 & 2000 \\
$m_{\tilde Q_{3}}$ & 2000 & 2000 & 2000 & 2500 & 500 \\
$A_{t}$ & 2300 & 2300 & 2300 & 2500 & 1000  \\\hline
$m_{\til_{1\dots 6}}$ & 489--515 & 738--765 & 984--1018 & 488--516 & 474--802   \\
$m_{\tinu_{1\dots 3}}$ & 496 & 747 & 998 & 496 & 496--797  \\
$m_{{\tilde\chi}_{1,2}^\pm}$ & 375--531 & 376--530 & 377--530  & 710--844 & 377--530   \\
$m_{{\tilde\chi}^0_{1\dots 4}}$ & 244--531 & 245--531 & 245--530 & 373--844 & 245--530   \\
$M_h$ & 126.6 & 127.0 & 127.3 & 123.8 & 123.1  \\
$M_H$ & 1300 & 1500 & 1799 & 1000 & 1700   \\
$M_{H^\pm}$ & 1302 & 1502 & 1801 & 1003 & 1701  \\
$m_{\tilde u_{1\dots 6}}$& 1909--2100 & 1909--2100 & 1908--2100 & 2423--2585 & 336--2000  \\
$m_{\tilde d_{1\dots 6}}$ & 1997--2004 & 1994--2007 & 1990--2011 & 2498--2503 & 474--2001  \\
$m_{\tilde g}$ & 2000 & 2000 & 2000 & 3000 & 2000  \\
\hline
\end{tabular}}
\caption{Selected points in the MSSM parameter space 
(upper part) and their corresponding spectra (lower part).  All 
dimensionful quantities are in $\gev$.}
\label{tab:spectra}
\end{table}

The specific values of these ten MSSM parameters are given in %
\refta{tab:spectra}. These are chosen to provide different patterns in the
various sparticle masses, but all lead to rather heavy spectra, that are
naturally, in agreement with the absence of SUSY signals at the LHC. In
particular, all points \ indicate the presence of rather heavy squarks and
gluinos above $1200\gev$ and heavy sleptons above $500\gev$ (where the LHC
limits would also permit substantially lighter sleptons). The values of $\MA$%
, $\tb$ and a large $A_{t}$ within intervals $(1000,1800)\gev$, $(10,50)\gev$
and $(1000,2500)\gev$ respectively, are fixed such that a light Higgs boson $%
h$ within the LHC-favoured range $(123,127)\gev$ is obtained.

Particularly, very low values of $M_{2}$ and $M_{3}$ are restricted by the
GUT relation and the absence of gluinos at the LHC, respectively. This is
reflected by our choice of $M_{2}$ and $\mu $ which makes gaugino masses
compatible with present LHC bounds. Furthermore, we require that all our
points lead to a prediction of the anomalous magnetic moment of the muon in
the MSSM, in order to fulfil the prevalent discrepancies between the
predictions of the Standard Model and the experimental values.

\subsection{Experimentally allowed values of $\deFABij$}

For selected reference scenarios some BPO are considered: BR($%
B^{0}\rightarrow \mu ^{+}\mu ^{-}),$ BR$(B\rightarrow X_{s}\gamma )$, and $%
\Delta M_{B_{s}}.$ The experimental values of these BPO are mentioned in %
\refta{tab:BPO-status}. Moreover, these experimental values allow one to put
bounds on flavour violating $\delta ^{\prime }s.$ For our analysis we took
MSSM parameters, BPO and their corresponding bounds on $\delta _{ij}^{FAB}$
from \cite{Arana-Catania:2014ooa}. The complete list of bounds are given in %
\refta{tab:boundsS1S6}. We have checked that the modified value of $M_A$ 
does not result in significant changes in the FCNC bounds reported 
in \cite{Arana-Catania:2014ooa} for LL and RR sectors,   however, there are some modifications in the
intervals of $\delta _{23}^{ULR}$\ and $\delta _{23}^{DLR}$ due to the vacuum stability constraints. More
stringent constraints on $\delta _{23}^{ULR}$\ and $\delta _{23}^{DLR}$\
come from vacuum stability conditions (mainly from CCB) as compared to FCNC
bounds. Bounds on $\delta
_{23}^{ULR}$\ and $\delta _{23}^{DLR}$\ from CCB and UFB are shown
in \refta{tab:ccb-ufb-bounds}. S5 is excluded for all $\delta
_{ij}^{FAB}$, except for $\delta _{23}^{DLR}$, because sizeable value of FV
deltas is not possible in S5 as it violates BPO constraints. One can
clearly differentiate from \refta{tab:boundsS1S6} and %
\refta{tab:ccb-ufb-bounds} that constraints on $\delta _{23}^{ULR}$\ and $%
\delta _{23}^{DLR}$\ coming from CCB and UFB are stronger than the FCNC
bounds.

\renewcommand{\arraystretch}{1.1} 
\begin{table}[]
\begin{center}
\begin{tabular}{|c|c|c|}
\hline
& Experimental Values & SM Predictions \\ \hline
$BR(B\rightarrow X_{s}\gamma)$ & $3.43 \pm0.22 \times 10^{-4}$ & $3.15 \pm
0.23 \times 10^{-4}$ \\ \hline
$BR(B^{0}\rightarrow \mu^{+}\mu^{-})$ & $3.0_{-0.9}^{+1.0}\times 10^{-9}$ & $%
3.23\pm 0.27\times 10^{-9}$ \\ \hline
$\Delta M_{B_{s}}$ & $116.4\pm 0.5\times 10^{-10} \mev$ & $%
117.1_{-16.4}^{+17.2}\times 10^{-10} \mev$ \\ \hline
\end{tabular}
\end{center}
\caption{Currents experimental status of BPO and SM predictions.}
\label{tab:BPO-status}
\end{table}
\renewcommand{\arraystretch}{1.55} 

\renewcommand{\arraystretch}{1.1}
\begin{table}[H]
\begin{center}
\resizebox{9.0cm}{!} {
\begin{tabular}{|c|c|c|} \hline
 & & Total allowed intervals \\ \hline
$\delta^{QLL}_{23}$ & \begin{tabular}{c}  S1 \\ S2 \\ S3 \\ S4 \\ S5 \end{tabular} &  
\begin{tabular}{c} 
(-0.27:0.28) \\ (-0.23:0.23) \\ (-0.12:0.06) (0.17:0.19)  \\ (-0.83:-0.78) (-0.14:0.14) \\ excluded \end{tabular} \\ \hline
$\delta^{ULR}_{23}$  & \begin{tabular}{c}  S1 \\ S2 \\ S3 \\ S4 \\ S5 \end{tabular}    
& \begin{tabular}{c} 
(-0.27:0.27) \\ (-0.27:0.27) \\ (-0.27:0.27) \\ (-0.22:0.22) \\ excluded   \end{tabular}   \\ \hline
$\delta^{DLR}_{23}$  & \begin{tabular}{c}  S1 \\ S2 \\ S3 \\ S4 \\ S5 \end{tabular}    & 
\begin{tabular}{c} 
(-0.0069:0.014) (0.12:0.13) \\ (-0.0069:0.014) (0.11:0.13) \\ (-0.0069:0.014) (0.11:0.13) \\  (-0.014:0.021) (0.17:0.19) \\ (0.076:0.12) (0.26:0.30)  \end{tabular}  \\ \hline
$\delta^{URL}_{23}$  & \begin{tabular}{c}  S1 \\ S2 \\ S3 \\ S4 \\ S5 \end{tabular}    & 
\begin{tabular}{c}
(-0.27:0.27) \\ (-0.27:0.27) \\ (-0.27:0.27) \\ (-0.22:0.22) \\ excluded  \end{tabular} 
  \\ \hline
$\delta^{DRL}_{23}$  & \begin{tabular}{c}  S1 \\ S2 \\ S3 \\ S4 \\ S5 \end{tabular}    &
\begin{tabular}{c} (-0.034:0.034) \\ (-0.034:0.034) \\ (-0.034:0.034) \\ (-0.062:0.062) \\ excluded  \end{tabular} 
  \\ \hline
$\delta^{URR}_{23}$ & \begin{tabular}{c}  S1 \\ S2 \\ S3 \\ S4 \\ S5 \end{tabular}   & \begin{tabular}{c} 
(-0.99:0.99) \\ (-0.99:0.99) \\ (-0.98:0.97) \\ (-0.99:0.99) \\ excluded  \end{tabular}    \\ \hline
$\delta^{DRR}_{23}$  & \begin{tabular}{c}  S1 \\ S2 \\ S3 \\ S4 \\ S5\end{tabular}    &
\begin{tabular}{c}  (-0.96:0.96) \\ (-0.96:0.96) \\ (-0.96:0.94) \\ (-0.97:0.97) \\ excluded  
\end{tabular}    \\ \hline
\end{tabular}}  
\end{center}
\caption{Present allowed (by BPO) intervals for the squark mixing parameters
  $\delta^{FAB}_{ij}$ for the selected S1-S5 MSSM points defined in \refta{tab:spectra}. 
}
\label{tab:boundsS1S6}
\end{table}
\renewcommand{\arraystretch}{1.55}

\renewcommand{\arraystretch}{1.1}
\begin{table}[H]
\begin{center}
\resizebox{9.0cm}{!} {
\begin{tabular}{|c|c|c|c|} \hline
 & & CCB Bounds & UFB Bounds\\ \hline
$\delta^{ULR}_{23}$  & \begin{tabular}{c}  S1 \\ S2 \\ S3 \\ S4 \\ S5 \end{tabular}    
& \begin{tabular}{c} 
-0.12:0.12\\
-0.12:0.12\\
-0.12:0.12\\
-0.09:0.09\\
-0.19:0.19

  \end{tabular}
& \begin{tabular}{c} 
-0.12:0.12\\
-0.13:0.13\\
-0.13:0.13\\
-0.09:0.09\\
-0.22:0.22

\end{tabular}   \\ \hline
$\delta^{DLR}_{23}$  & \begin{tabular}{c}  S1 \\ S2 \\ S3 \\ S4 \\ S5 \end{tabular}    & 
\begin{tabular}{c} 
-0.003:0.003\\
-0.003:0.003\\
-0.003:0.003\\
-0.002:0.002\\
-0.006:0.006

\end{tabular} 
& \begin{tabular}{c} 
-0.003:0.003\\
-0.003:0.003\\
-0.003:0.003\\
-0.002:0.002\\
-0.005:0.005

\end{tabular} \\ \hline
\end{tabular}}  
\end{center}
\caption{Constraints on $\delta^{FAB}_{ij}$ originating from vacuum stability condition for the selected S1-S5 MSSM points defined in \refta{tab:spectra}. 
}
\label{tab:ccb-ufb-bounds}
\end{table}
\renewcommand{\arraystretch}{1.55}

\subsection{$\Gamma(\tilde{g} \rightarrow \tilde{q}_i \bar{q}_i)$}

We have calculated tree level partial decay widths of gluino into quarks and
the lightest squarks. In our analysis, we use MSSM input parameters and corresponding
physical mass spectra of squarks and gluino given in \refta{tab:spectra}.
This mass range dictates that kinematically allowed decays are $\tilde{g}%
\rightarrow q\tilde{u}_{1},$\ $\tilde{g}\rightarrow q\tilde{d}_{1}$\ (only
1st generation squarks) for S1 to S3, whereas, for S4 and S5 all three
generations squarks can be accommodated i.e. $\tilde{g}\rightarrow q\tilde{u}%
_{i},$\ $\tilde{g}\rightarrow q\tilde{d}_{i}$\ with $i=1,...6$.

Conversely, in this paper $\tilde{g}\rightarrow q\tilde{u}_{1},$\ $\tilde{g}%
\rightarrow q\tilde{d}_{1}$\ decay modes will be our prime focus. The
dependence of partial decay width on QFV parameters of the MSSM have been
analyzed within the allowed experimental ranges, as given in \refta
{tab:boundsS1S6} and \refta{tab:ccb-ufb-bounds}. In our analysis, we considered
 tree level partial decay widths only because loop level corrections for LL and RR mixings are
 found to be small \cite{Eberl:2017pbu}. The mixing in the LR and RL sectors 
 was ignored even at the tree level in \cite{Eberl:2017pbu} due to the strong constraints on 
 these sectors from vacuum stablity conditions. However, we are able to show that the LR and RL 
 mixing contributions can be significant for some part of the parameter space.

\begin{figure}[ht!]
\begin{center}
\psfig{file=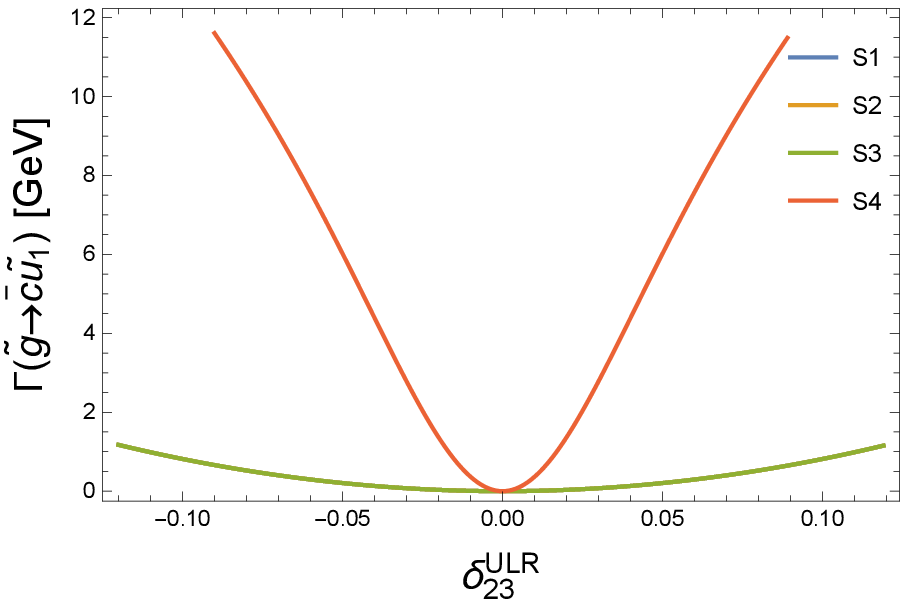  ,scale=0.85,angle=0,clip=} %
\psfig{file=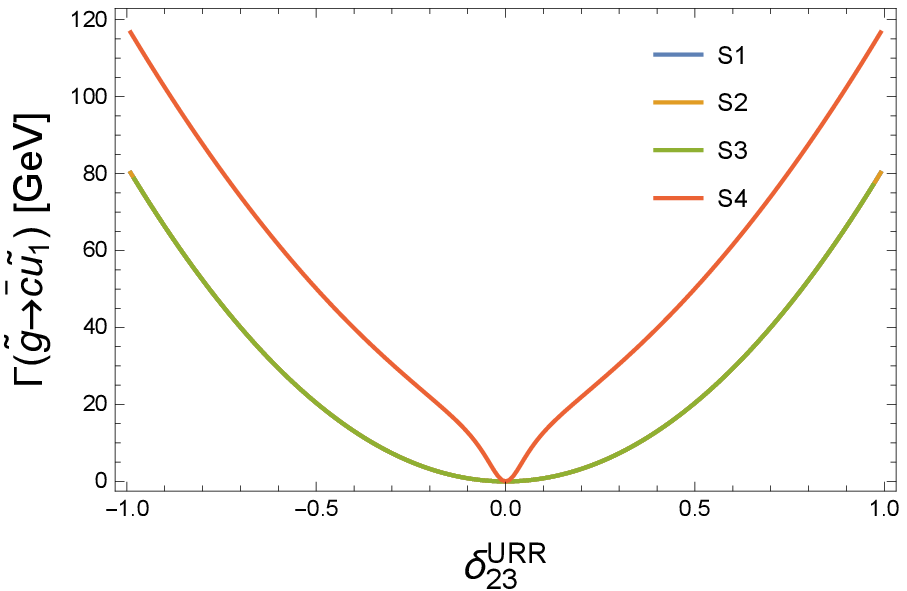  ,scale=0.85,angle=0,clip=}\\[0pt]
\end{center}
\caption{Partial decay width of $\tilde{g}\rightarrow \bar{c}\tilde{u}_{1}$
mode as a function of $\protect \delta_{23}^{ULR}$ (left) and $\protect \delta%
_{23}^{URR}$ (right).}
\label{fig1}
\end{figure}

In \reffi{fig1}, we have analysed the dependence of QFV parameters on
partial decay width of $\tilde{g}\rightarrow \bar{c}\tilde{u}_{1}$. As, $%
\tilde{u}_{1}$\ is mainly an up-type squark thereby only up-type QFV
parameters are relevant for the said analysis. The contribution of $\delta
_{23}^{QLL}$\ is found to be negligible and is not shown here. In the left plot of \reffi{fig1}, we show 
$\tilde{g}\rightarrow \bar{c}\tilde{u}_{1}$ as a function of $\delta_{23}^{ULR}$. 
In spite of the stringent constraints on this delta coming from CCB, the QFV 
contribution can reach upto 11.6 GeV for scenario S4 as indicated by the red line.
For the three scenarios: S1, S2 and S3, the behavior is the same and is shown collectively by the
green line. 
In \reffi{fig1}(right), we show the dependence of $\Gamma(\tilde{g}\rightarrow \bar{c}\tilde{u}_{1})$ on $\delta _{23}^{URR}$. The entire region $(-0.99:0.99)$ is allowed for $\delta _{23}^{URR}$. In the first 2 scenarios, the behaviour is the same i.e. 80 GeV (shown by
green) and for S3 $\Gamma(\tilde{g}\rightarrow \bar{c}\tilde{u}_{1})$ reaches 78 GeV, but for S4, the partial decay width goes upto 116.7 $\gev$ as shown by the red line.


\begin{figure}[ht!]
\begin{center}
\psfig{file=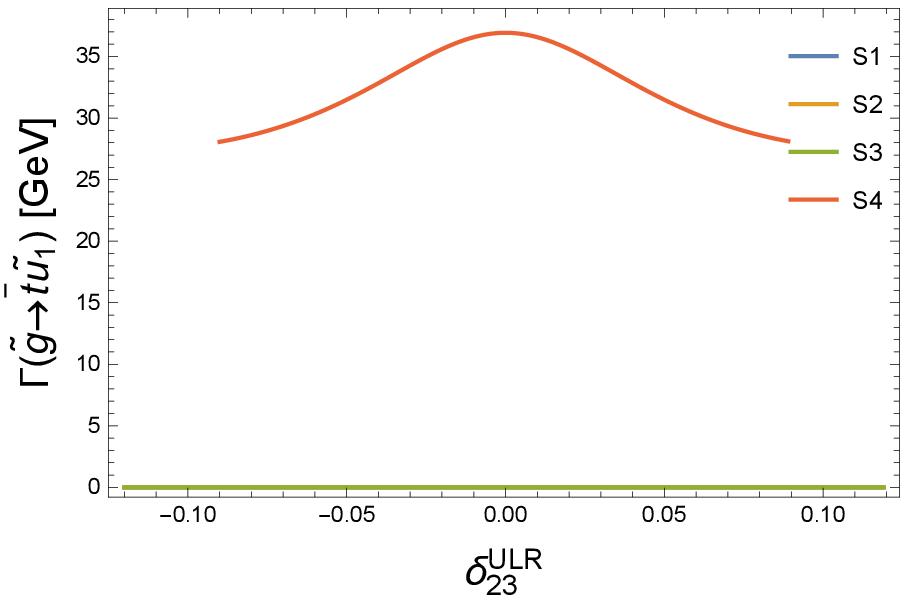  ,scale=0.85,angle=0,clip=} %
\psfig{file=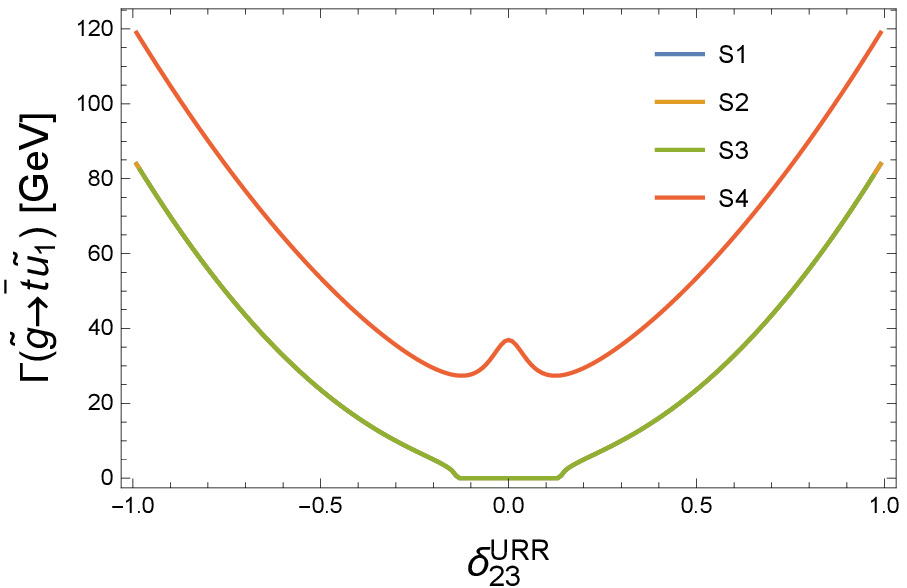  ,scale=0.85,angle=0,clip=}\\[0pt]
\end{center}
\caption{Partial decay width of $\tilde{g}\rightarrow \bar{t}\tilde{u}_{1}$
mode as a function of $\protect \delta_{23}^{ULR}$ (left) and $\protect \delta%
_{23}^{URR}$ (right).}
\label{fig2}
\end{figure}

\ In \reffi{fig2} we show the partial decay width of $\tilde{g}\rightarrow \bar{t}\tilde{u}_{1}$ 
as a function of QFV parameters $\delta^{ULR}_{23}$ (left) and $\delta^{URR}_{23}$ (right). 
For the first 3 scenarios, $\delta^{ULR}_{23}$ gives negligible contribution to the partial decay
width while for S4, its contribution reaches upto 8 $\gev$\ for the
region $(-0.14:0.14)$ as can be seen in the left plot of \reffi{fig2}. We would like
to point however that for the point S4, a small interval (-0.83:-0.76) for $\delta _{23}^{QLL}$
is allowed where the partial decay width can reach upto $\approx 36 \gev$. On the other hand, the right plot in \reffi{fig2}
represent the contributions resulting from $\delta^{URR}_{23}$. Here again almost entire interval is allowed for $\delta^{URR}_{23}$ and its contribution to the partial decay width $\Gamma(\tilde{g}\rightarrow \bar{t}\tilde{u}_{1})$ amounts to $84 \gev$ for first 2 scenarios and for S3, S4 decay width is $82 \gev$, $118.9 \gev$, respectively.

\begin{figure}[ht!]
\begin{center}
\psfig{file=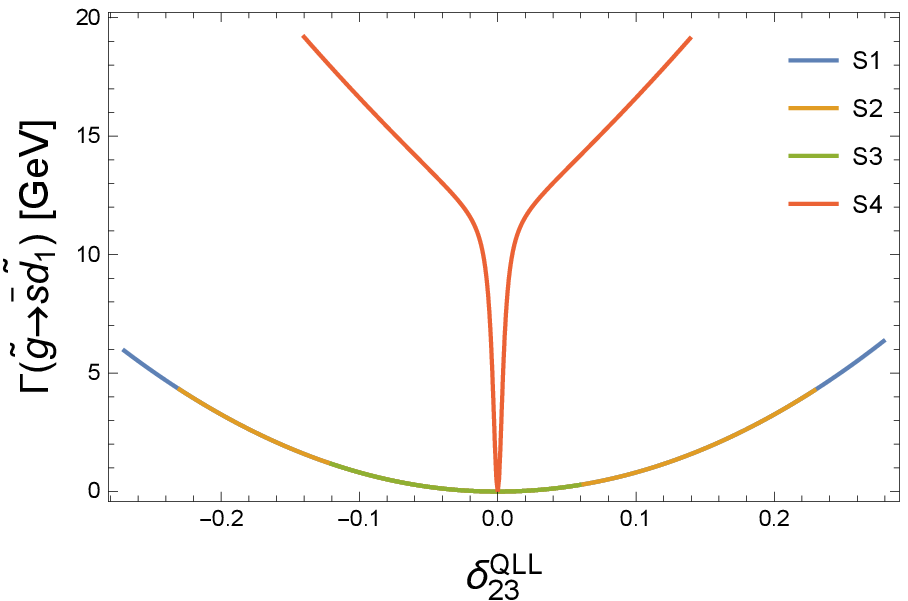  ,scale=0.75,angle=0,clip=} %
\psfig{file=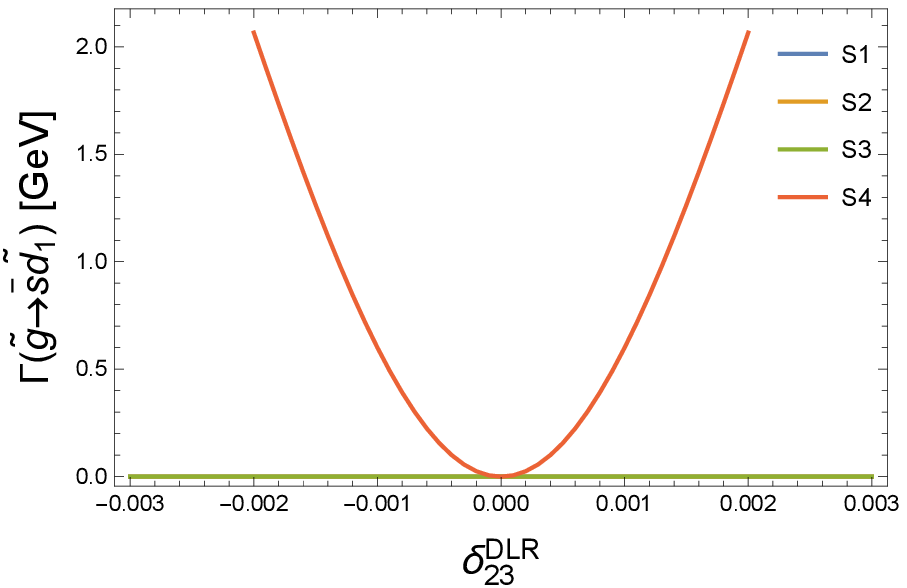  ,scale=0.75,angle=0,clip=}\\[0pt]
\psfig{file=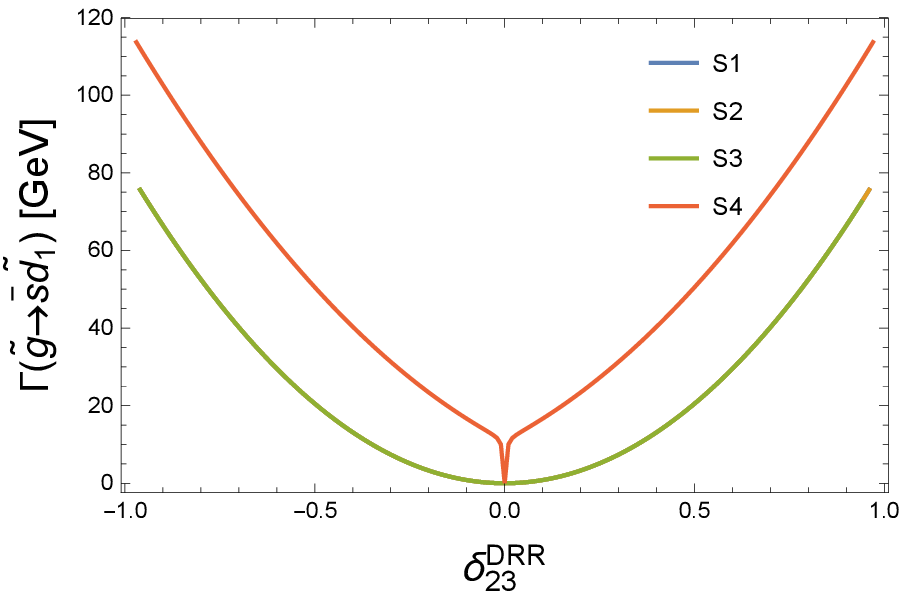 ,scale=0.75,angle=0,clip=}
\end{center}
\caption{Partial decay width of $\tilde{g}\rightarrow \bar{s}\tilde{d}_{1}$
mode as a function of $\protect \delta_{23}^{QLL}$ (upper left), $\protect%
\delta_{23}^{DLR}$ (upper right) and $\protect \delta _{23}^{DRR}$ (lower
plot).}
\label{fig3}
\end{figure}

\begin{figure}[ht!]
\begin{center}
\psfig{file=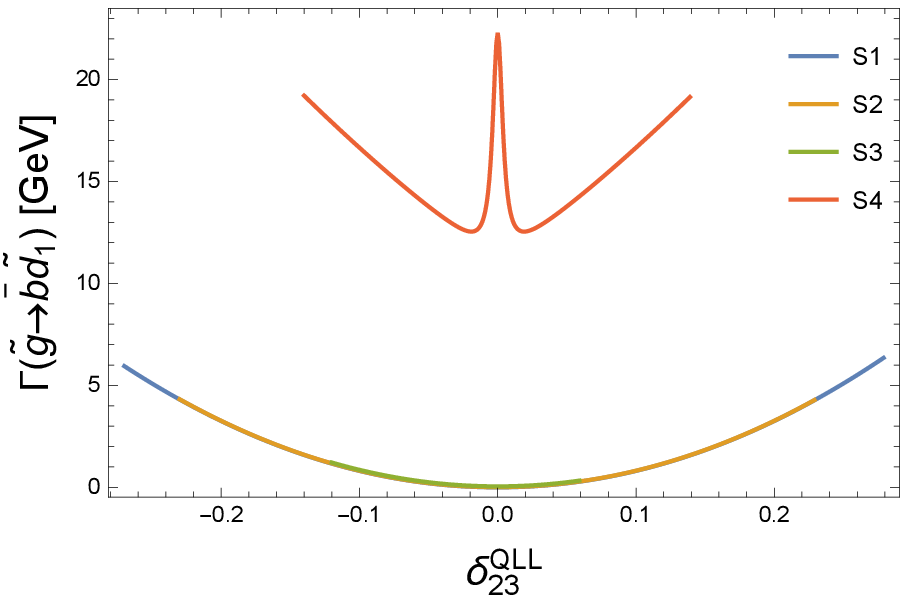  ,scale=0.75,angle=0,clip=} %
\psfig{file=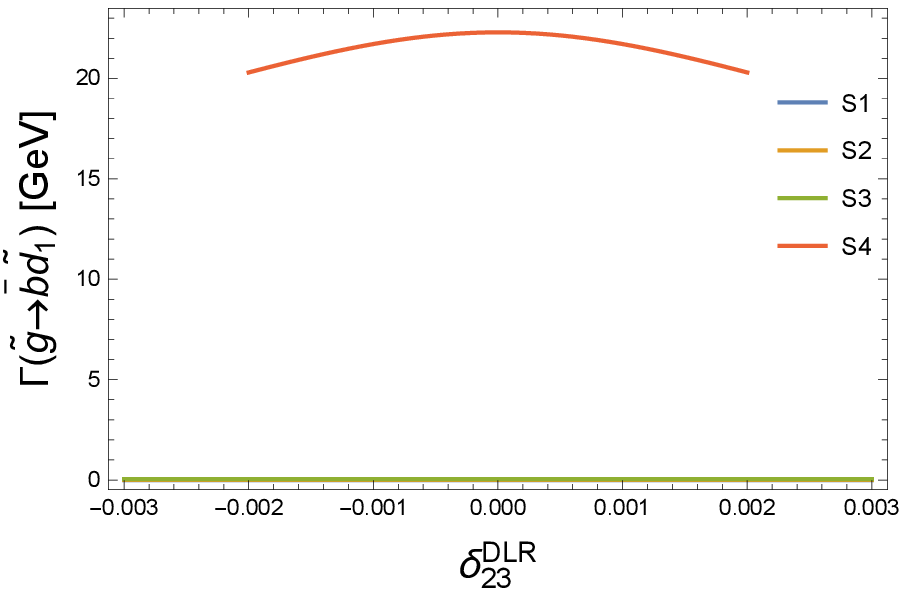  ,scale=0.75,angle=0,clip=}\\[0pt]
\psfig{file=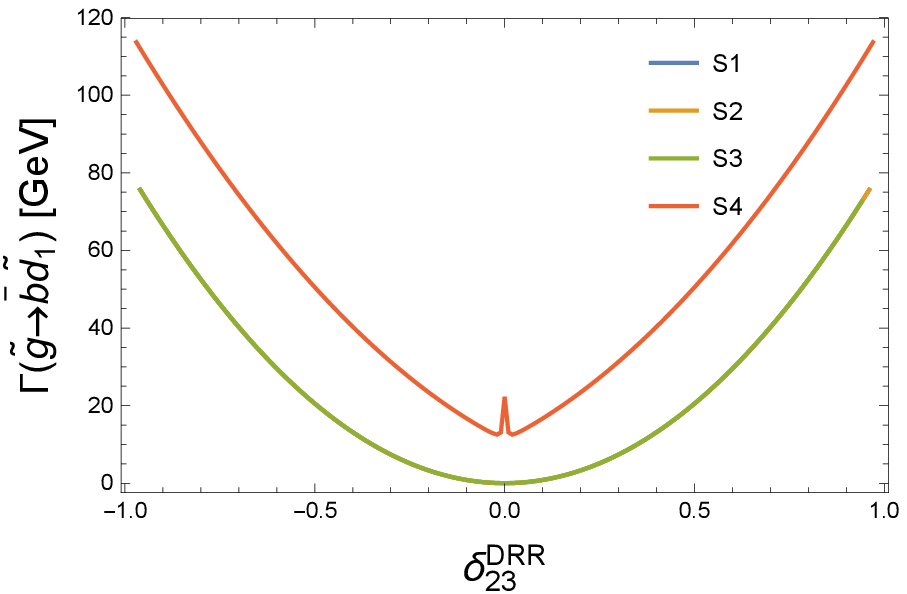 ,scale=0.75,angle=0,clip=}
\end{center}
\caption{Partial decay width of $\tilde{g}\rightarrow \bar{b}\tilde{d}_{1}$
mode as a function of $\protect \delta_{23}^{QLL}$ (upper left), $\protect%
\delta_{23}^{DLR}$ (upper right) and $\protect \delta _{23}^{DRR}$ (lower
plot).}
\label{fig4}
\end{figure}

In \reffis{fig3} and \ref{fig4}, we show the decay width $\Gamma (\tilde{g}\rightarrow \bar{s}\tilde{d}_{1})$
and $\Gamma (\tilde{g}\rightarrow \bar{b}\tilde{d}_{1})$ respectively, as a function of QFV parameters. 
The effect of $\delta^{QLL}_{23}$ on $\Gamma (\tilde{g}\rightarrow \bar{s}\tilde{d}_{1})$
is significant and is shown in the upper left plot of \reffi{fig3} where it can reach upto 
$6 \gev$, $4 \gev$ and $1 \gev$  for the S1, S2 and S3, respectively. While it can reach upto $19 \gev$ in S4 for the interval (-0.14:0.14) and upto $92 \gev$ for the interval (-0.83:-0.76).  In the upper right plot of \reffi{fig3}, 
we show the dependence of $\Gamma (\tilde{g}\rightarrow \bar{s}\tilde{d}_{1})$ on $\delta^{DLR}_{23}$. 
Here the UFB/CCB constraints on $\delta _{23}^{DLR}$ are very stringent and only a small interval (-0.002,0.002)
is allowed for the point S4. In this small interval, the $\Gamma (\tilde{g}\rightarrow \bar{s}\tilde{d}_{1})$ is $2 \gev$. However for other scenarios, the effect of $\delta^{DLR}_{23}$ 
on the $\Gamma (\tilde{g}\rightarrow \bar{s}\tilde{d}_{1})$ remains negligible.
The effects of $\delta _{23}^{DRR}$ are shown in \reffi{fig3} (lower center plot). 
Its contribution in S1, S2, S3 can amount to $75.6 \gev$ while for the point S4, 
the partial decay width reaches upto $114 \gev$. 

In the upper left plot of \reffi{fig4}, we have analyzed the $\Gamma (\tilde{g}\rightarrow \bar{b}\tilde{d}_{1})$ as a
function of $\delta _{23}^{QLL}$. In the first scenario, the contribution of $\delta^{QLL}_{23}$ 
can result in the increase of $\Gamma (\tilde{g}\rightarrow \bar{b}\tilde{d}_{1})$ upto $6\gev$ under the allowed interval. For S2 and S3 the contribution is $4 \gev$ and $1 \gev$, respectively.
For the point S4, $\delta^{QLL}_{23}$ gives negative contribution to the $\Gamma (\tilde{g}\rightarrow \bar{b}\tilde{d}_{1})$ 
in the interval (-0.01,0.01). However for $\delta^{QLL}_{23}>0.01$, the $\Gamma (\tilde{g}\rightarrow \bar{b}\tilde{d}_{1})$ 
gets positive contributions. The overall effect can reach upto 10 $\gev$ depending upon the value of $\delta^{QLL}_{23}$. However, for the interval (-0.83:-0.76), the $\delta^{QLL}_{23}$ gives $\Gamma (\tilde{g}\rightarrow \bar{b}\tilde{d}_{1})= 92 \gev$.  
In \reffi{fig4} (upper left plot), we can see the $\Gamma (\tilde{g}\rightarrow \bar{b}\tilde{d}_{1})$ as a function $\delta^{DLR}_{23}$.
Due to the stringent constraints on this parameter, its contribution is negligible except in the scenario S4 where it can be as high as 
$20 \gev$. As a last step, we show $\Gamma (\tilde{g}\rightarrow \bar{b}\tilde{d}_{1})$ as a function of $\delta^{DRR}_{23}$. 
The contribution goes upto $76\gev$ for the first three scenarios and upto $114 \gev$ for the scenario S4.
We summarize our findings in the \refta{tab:summarytable} where we show the flavor conserving and FV contributions to the 
partial decay width of different decay modes of gluino. 

To summarize, in spite of strong constraint on QFV parameters coming from BPO and vacuum stability condition,
QFV  gluinos  decays do not only get contributions from LL/RR mixing but LR/RL mixings can also make colossal
contributions to the partial decay width. These contributions may have important influence on the experimental 
searches for gluinos at LHC and future colliders. 

\begin{table}[h!]
\begin{center}
\begin{tabular}
[c]{|c|c|c|c|c|}\hline
\multicolumn{5}{|c|}{$\tilde{g}\rightarrow\bar{s}\tilde{d}_{1}$}\\\hline
Point & No FV & $\delta_{23}^{QLL}$ & $\delta_{23}^{DLR}$ & $\delta_{23}%
^{DRR}$\\\hline
$S1$ & $0$ & $6.39$ & $0.61\times10^{-3}$ & $75.6$\\\hline
$S2$ & $0$ & $4.3$ & $0.61\times10^{-3}$ & $75.6$\\\hline
$S3$ & $0$ & $1.16$ & $0.61\times10^{-3}$ & $75.6$\\\hline
$S4$ & $0$ & $19.18$ & $2.06$ & $113.6$\\\hline
\multicolumn{5}{|c|}{$\tilde{g}\rightarrow\bar{b}\tilde{d}_{1}$}\\\hline
Point & No FV & $\delta_{23}^{QLL}$ & $\delta_{23}^{DLR}$ & $\delta_{23}%
^{DRR}$\\\hline
$S1$ & $0$ & $6.35$ & $0.16\times10^{-2}$ & $75.61$\\\hline
$S2$ & $0$ & $4.32$ & $0.10\times10^{-1}$ & $75.61$\\\hline
$S3$ & $0$ & $1.22$ & $0.50\times10^{-1}$ & $75.61$\\\hline
$S4$ & $22.29$ & $19.21$ & $20.29$ & $113.7$\\\hline
\multicolumn{5}{|c|}{$\tilde{g}\rightarrow\bar{c}\tilde{u}_{1}$}\\\hline
Point & No FV & $\delta_{23}^{QLL}$ & $\delta_{23}^{ULR}$ & $\delta_{23}%
^{URR}$\\\hline
$S1$ & 0 & 0 & 1.117 & 80.2\\\hline
$S2$ & 0 & 0 & 1.17 & 80.2\\\hline
$S3$ & 0 & 0 & 1.17 & 78.57\\\hline
$S4$ & 0 & 0 & 11.6 & 116.68\\\hline
\multicolumn{5}{|c|}{$\tilde{g}\rightarrow\bar{t}\tilde{u}_{1}$}\\\hline
Point & No FV & $\delta_{23}^{QLL}$ & $\delta_{23}^{ULR}$ & $\delta_{23}%
^{URR}$\\\hline
$S1$ & 0 & 0 & 0 & 83.9\\\hline
$S2$ & 0 & 0 & 0 & 83.9\\\hline
$S3$ & 0 & 0 & 0 & 82.3\\\hline
$S4$ & 36.9 & 0 & 28.06 & 119\\\hline
\end{tabular}
\caption{Partial decay width of $\tilde{g}\rightarrow\bar{s}\tilde{d}_{1}$, 
$\tilde{g}\rightarrow\bar{b}\tilde{d}_{1}$, $\tilde{g}\rightarrow\bar{c}\tilde{u}_{1}$ and $\tilde{g}\rightarrow\bar{t}\tilde{u}_{1}$ with and without flavor violation for the selected parameter points shown in \refta{tab:spectra}.}
\label{tab:summarytable}
\end{center}
\end{table}


\clearpage

\section{Conclusions}
\label{sec:conclusions}
Supersymmetry (SUSY), in spite of being one of the best candidate beyond the Standard Model (SM), 
still remains undetected. For SUSY searches at the LHC and future colliders, it is important to study sparticle decays,
particularly the decays of the strongly interacting SUSY particles like squarks and gluinos. 
On the other hand, limits on the sparticle masses are getting higher and higher with each passing day. 
For example, gluino masses $\leq $ 1900
$\gev$ are excluded ~\cite{Aad:2020nyj}. It is therefore, important to study gluino decays with high precison.  
In this paper we have investigated the effect of squark flavor mixing, parameterized in terms of $\delta _{ij}^{FAB}$
parameters, on the quark flavour violating decays of gluinos into lightest
squarks $(\tilde{g}\rightarrow \overline{c}\tilde{u}_{1},$ $\tilde{g}%
\rightarrow \overline{t}\tilde{u}_{1},$ $\tilde{g}\rightarrow \overline{s}%
\tilde{d}_{1},$ $\tilde{g}\rightarrow \overline{b}\tilde{d}_{1})$. 

We have analyzed the effect of squark mixing in the LL, LR/RL and RR part of the squarks mass matrices on
partial decay width of $\tilde{g}\rightarrow q\tilde{q}$. We choose four reference scenarios (with a slight modification 
in the $M_A$), first studied in \cite{Arana-Catania:2014ooa}, with the corresponding constraints on the flavor violating (FV) 
deltas coming from B Physics Observables (BPO). 
For the said scenerios, we have calculated the constraints on $\delta _{ij}^{FLR,FRL}$,
using charge and color breaking minima (CCB) and unbounded from below (UFB) minima. It is thereby observed that the
constraints from CCB and UFB are more stringent than the ones obtained via
BPO. While it is true that the mixng in the RR part of the squarks mass matrices gives the highest contributions 
ranging from $75-120 \gev$, we find however that the mixing in the LL and LR/RL sector can also be important. 
For instance, for $\delta^{QLL}_{23}$, the $\Gamma(\tilde{g}\rightarrow \overline{s}\tilde{d}_{1})= 92 \gev$ for some parameter points.
Similarly $\delta^{ULR}_{23}$ can contribute $\approx 12 \gev$ to $\Gamma(\tilde{g}\rightarrow \overline{c}\tilde{u}_{1})$ and $\approx 8 \gev$ to $\Gamma(\tilde{g}\rightarrow \overline{t}\tilde{u}_{1})$. This analysis shows the importance of 
QFV parameters which could have an important influence on the search for gluinos and the determination of the MSSM
parameters at HL-LHC or HE-LHC.\vspace{-0.5em}
\subsection*{Acknowledgments}
We would like to thank Sven Heinemeyer and Mario E. Gomez for their helpful suggestions during the preparation of this manuscript.



\bibliography{gdecay}{}
\bibliographystyle{unsrt}

\end{document}